\documentstyle[prl,aps,epsfig,preprint,floats]{revtex}
\begin{document}
\tighten
\def\lsi{\raise0.3ex\hbox{$<$\kern-0.75em\raise-1.1ex\hbox{$\sim$}}}
\def\gsi{\raise0.3ex\hbox{$>$\kern-0.75em\raise-1.1ex\hbox{$\sim$}}}
\newcommand{\lsim}{\mathop{\lsi}}
\newcommand{\gsim}{\mathop{\gsi}}
\def\tphi{\tilde{\phi}}
\def\tE{\tilde{E}}
\def\xi{x\!+\!\hat{\imath}}
\def\xj{x\!+\!\hat{\jmath}}
\def\xk{x\!+\!\hat{k}}

%@twocolumnfalse\endcsname

\title{Electroweak preheating on a lattice}
\author{
A. Rajantie,$^{1,3}$\thanks{E-mail: a.k.rajantie@damtp.cam.ac.uk}
P.M. Saffin$^{1,2}$\thanks{E-mail: P.M.Saffin@durham.ac.uk}
and
E.J. Copeland$^3$\thanks{E-mail: e.j.copeland@sussex.ac.uk}
}
\address{
$^1$DAMTP, CMS, University of Cambridge,
Wilberforce Rd, Cambridge, CB3 0WA,~~~U.~K.\\
$^2$Centre for Particle Theory, Durham University, South Road, Durham,
DH1 3LE,~~~U.~K.
$^3$Centre for Theoretical Physics,
University of Sussex, Falmer, Brighton BN1 9QJ,~~~U.~K.\\
}

%\date{\today}
\date{May 25, 2001}
\maketitle
\begin{abstract}
In many inflationary models, 
a large amount of energy is transferred rapidly to the
long-wavelength matter fields during a period of preheating after
inflation. We study how this changes the dynamics of the electroweak
phase transition if inflation ends at the
electroweak scale. 
We simulate a 
classical SU(2)$\times$U(1)+Higgs model with initial conditions in
which
the energy is concentrated in the long-wavelength Higgs modes.
With a suitable initial energy density, the electroweak
symmetry is restored non-thermally but broken again when
the fields thermalize. During this symmetry restoration, baryon number
is violated, and we measure 
its time evolution, pointing out that it is highly non-Brownian.
This makes it difficult to estimate the generated baryon asymmetry.
\end{abstract}
\pacs{PACS: 11.15.Ha, 11.15.Kc, 98.80.Cq\\
DAMTP-2000-134, DTP/00/105, SUSX-TH-00-022
 \hfill
hep-ph/0012097
}
%\vskip2pc]
%%%%%%%%%%%%%%%%%%%%%%%%%%%%%%%%%%%%%%%%%%%%%%%%%%%%%%

\section{Introduction}

In order to explain the observed baryon asymmetry of the universe, a
theory must satisfy three conditions \cite{Sakharov:1967dj}; baryon number
violation, C and CP violation and deviation from thermal equilibrium.
Grand unified theories offer the required properties, but in
addition to baryons, they also lead to formation of magnetic
monopoles. This monopole problem can be solved by a subsequent period
of inflation, but it would also wipe out any baryon asymmetry.

Although the baryon number is conserved perturbatively
in the electroweak theory, there are non-perturbative effects that
violate it. At zero temperature, this effect is negligible, because
the baryon number can only be changed by strongly suppressed tunneling
processes\cite{tHooft76}. However, there are classical field configurations,
sphalerons \cite{Manton85}, that interpolate between different baryon
numbers and become unsuppressed
at high temperatures \cite{kuzmin85}.
Thus, provided that CP is violated, the electroweak theory can explain
the baryon asymmetry by itself, as long as the electroweak phase
transition gives the necessary non-equilibrium state.

However, sphaleron processes do not stop immediately after the
electroweak transition, and thus
the baryon asymmetry can be washed out by the sphalerons, unless
the electroweak symmetry is already strongly broken when the universe
returns to thermal equilibrium. In the standard big bang scenario, only
a strongly first-order phase transition can prevent this baryon
washout, 
but lattice Monte
Carlo simulations have shown that the thermal transition cannot be strong
enough in the standard model \cite{Kajantie:1997qd}. Although it is possible
in some extensions, such as MSSM \cite{Laine:2001rm}, this has led people to
consider other possibilities.

A particularly attractive idea, suggested in
Refs.~\cite{Krauss:1999ng,Garcia-Bellido:1999sv}, is based on
the non-equilibrium dynamics at the end of inflation.
When inflation ends the universe is very cold and all the energy
is stored in the inflaton field, which subsequently decays into
the standard model fields, thus heating up the universe to the
reheat temperature $T_{\rm rh}$.
Instead of ordinary perturbative decays this can take place
non-perturbatively, and much faster, if the inflaton starts to
resonate with the matter fields \cite{kofman94}. In this period of
parametric resonance, called preheating, the energy is transferred
mainly to the modes with longest wavelengths, and they reach
an effective equilibrium state at a temperature $T_{\rm eff}$ which
can be much higher than $T_{\rm rh}$.
This makes it possible to effectively restore symmetries that are
spontaneously broken at $T_{\rm rh}$ \cite{kofman96,Tkachev:1996md}.
Eventually, the universe thermalizes and cools down to the equilibrium
temperature $T_{\rm rh}$, undergoing a non-thermal phase transition
to the broken phase.

In the special case in which $T_{\rm rh}$ is below the electroweak
critical temperature, baryon number is violated by sphalerons and the
system is far from equilibrium during the
period of non-thermal symmetry
restoration, and therefore the conditions for baryogenesis are
satisfied~\cite{Krauss:1999ng,Garcia-Bellido:1999sv}.
The washout problem is avoided if $T_{\rm rh}$ is low enough, because
the rate at which the universe cools is determined by the decay rate
of gauge bosons, and is much faster than the decay rate of baryons.

Because of the non-perturbative, non-equilibrium nature of the
process, the only reliable way of studying it is to simulate it
numerically.
Also, the fluctuations produced during preheating
have large occupation numbers so they
can be considered as
interacting classical waves~\cite{khlebnikov96,Prokopec:1997rr}.
Thus the dynamics of the system can be studied simply by solving
the classical equations of motion
numerically, which is a relatively straightforward task.

Non-thermal symmetry restoration was first observed in numerical
simulations in a scalar field model in Ref.~\cite{khlebnikov98}.
More recently, two of us studied the same process in the Abelian
Higgs model~\cite{rajantie00}, which has a very similar structure
to the electroweak theory. The results of the simulations support
the scenario, but unfortunately there is no analogue to the
Chern-Simons number in the three-dimensional Abelian Higgs model, and
thus the question of baryogenesis could not be addressed directly.
The generated baryon asymmetry has been measured in one-dimensional
Abelian simulations \cite{Garcia-Bellido:1999sv}, but it is not
possible to convert the results into three dimensions, and
therefore a quantitative test of the scenario is still missing.

In this paper, we study the dynamics of
the full electroweak theory
in non-equilibrium conditions. The methods we develop
can be applied to studies of baryogenesis during
preheating, in a given inflationary model. However, such simulations
would necessarily be very model-dependent and therefore we consider
only a simplified scenario in this paper.
In that scenario, we find that the rate of true topological sphaleron
transitions is slower than expected because of the interaction between
the gauge and Higgs fields.

We start by
reviewing the scenario of electroweak baryogenesis at preheating
in Sect.~\ref{sect:baryogenesis}. We only simulate the CP
invariant case, but since CP violation is necessary for
baryogenesis, we discuss the possibilities of including it in the
simulations in Sect.~\ref{sect:cp}. In Sect.~\ref{sect:approx}, we
discuss the approximations that are needed in order to make the
simulations feasible and show that all of them are justified.
Numerical simulations also require that the space-time is
discretized, and we present the details of this in
Sect.~\ref{sect:lattice}. In Sect.~\ref{sect:params}, we discuss
how the requirement that the baryon number is not washed out when
the system thermalizes, restricts the parameters of the theory. In
Sects.~\ref{sect:symmresto} and \ref{sect:Ncs} we present the
results of our simulations, and in Sect.~\ref{sect:concl}, we
discuss the conclusions we can draw from them. Some technical
details about the simulations are presented in the appendices.

\section{Baryogenesis at preheating}
\label{sect:baryogenesis}
The scenario of electroweak baryogenesis at
preheating~\cite{Krauss:1999ng,Garcia-Bellido:1999sv}
assumes a model of inflation in which the energy scale of
the inflation is so low
that the equilibrium temperature of the universe never
exceeds the electroweak critical
temperature after inflation.
The inflaton is assumed to couple to the standard model Higgs field.

The inflation dilutes the inhomogeneities in all the fields away and results
in a cold, vacuum configuration. However, the zero mode of the inflaton
has a large value, and
after inflation ends it oscillates with some frequency
$\omega$. This starts a parametric resonance with those Fourier
modes of the Higgs field that have the same frequency\cite{kofman94}.
This resonance amplifies the quantum fluctuations
of the Higgs field exponentially and transfers very rapidly
a significant fraction of the energy of the
inflaton to the Higgs field. If the frequency $\omega$ is small enough,
the resonance only excites
the long-wavelength part of the Higgs power spectrum,
but the occupation numbers of those modes becomes huge.

The energy subsequently
flows into the gauge bosons because of their strong
coupling to the Higgs field, and soon a quasi-equilibrium state
is reached in which the bosonic modes below a certain cutoff $k_*$
are approximately in thermal equilibrium while the other modes
are still in vacuum \cite{Prokopec:1997rr}.
Since the energy density is concentrated in a small number of degrees
of freedom,
the effective temperature $T_{\rm eff}$
of the long-wavelength modes is much higher than the reheat
temperature $T_{\rm rh}$,
and therefore the electroweak symmetry can be
restored~\cite{kofman96,Tkachev:1996md,rajantie00}.

We can estimate that $k_*\sim\alpha_WT_{\rm eff}
$~\cite{rajantie00},
because above that scale
the interactions are perturbative and
therefore the thermalization rate of the modes
with $k\gsim \alpha_WT_{\rm eff}$
is suppressed by powers of coupling constants.
It is important that the scale $k_*$ is the same as the typical
size of a sphaleron configuration~\cite{Garcia-Bellido:1999sv}.
Thus the sphaleron
transitions
are insensitive to the physics at higher momenta, and we can try and 
estimate
their rate in the same way as in thermal equilibrium.
In the broken phase, the sphaleron rate $\Gamma_{\rm sph}$,
defined as the number of
sphaleron processes per unit volume per unit time,
is exponentially suppressed by the Higgs expectation value,
but in the symmetric phase it is unsuppressed.
In the presence of an
ultraviolet cutoff $k_*$ at temperature $T_{\rm eff}$,
the classical sphaleron rate $\Gamma_{\rm sph}$ in the symmetric
phase
has been estimated to be~\cite{arnold}
\begin{equation}
\label{equ:sphest}
\Gamma_{\rm sph}\approx \alpha_W^5T_{\rm eff}^5/k_*\sim
\alpha_W^4T_{\rm eff}^4.
\end{equation}
Each sphaleron process changes the SU(2) Chern-Simons number
\begin{equation}
N_{\rm CS}(t) - N_{\rm CS}(0) = {g^2 \over 32 \pi^2} \int_0^t dt\int d^3 x
\epsilon^{\mu\nu\rho\sigma} {1 \over 2} W^a_{\mu\nu} W^a_{\rho\sigma}
\label{cs_number}
\end{equation}
by one, leaving the U(1) equivalent,
\begin{equation}
n_{\rm CS}(t) - n_{\rm CS}(0) = {g'^2 \over 32 \pi^2} \int_0^t dt\int d^3 x
\epsilon^{\mu\nu\rho\sigma} {1 \over 2} B_{\mu\nu} B_{\rho\sigma}
\label{ncs_number}
\end{equation}
unchanged.
Therefore, $N_{\rm CS}$ changes wildly during the non-thermal
symmetry restoration, and
the baryon current is not conserved, due to
the anomaly,
\begin{eqnarray}
\partial_{\mu} J^{\mu}_{B} &=&\frac{1}{32\pi^2}\epsilon^{\mu\nu\rho\sigma}
\left[ \frac{g^2}{2}W^a_{\mu\nu}W^a_{\rho\sigma}
      +\frac{g'^2}{2}B_{\mu\nu}B_{\rho\sigma}\right].
\end{eqnarray}
The system is strongly out of equilibrium and
therefore baryon asymmetry can be generated provided that
CP violation is strong enough.
As was pointed out in Ref.~\cite{Garcia-Bellido:2000px},
the oscillations of
the inflaton can enhance this process.

The effective temperature $T_{\rm eff}$ decreases gradually,
partly because the bosonic modes with higher momenta become
excited but mainly because of decays of the bosons into fermions.
The experimental value for the decay rate, $\gamma\approx 2$~GeV,
is much higher than the rate of baryon number violation
\begin{equation}
\label{equ:gammab}
\gamma_B\approx \frac{39}{4}\frac{\Gamma_{\rm sph}}{T^3_{\rm eff}},
\end{equation}
which implies that the sphaleron processes stop instantaneously
when the system enters the Higgs phase, and the baryon asymmetry
generated during the period of symmetry restoration is not washed
out. Simulations in a 1+1D toy model support this
scenario\cite{Garcia-Bellido:1999sv}, but obviously they cannot
decide whether it can really work in the context of the realistic
standard model. Our aim in this paper is to partially address that
question.

We are not going to address the issue of finding a model of
inflation that would have a low enough reheat temperature and
would produce the necessary initial conditions. In fact, we don't
treat the inflaton as a dynamical field, but simply assume that
the state of the system after the parametric resonance has
transferred the energy to the Higgs field, can be approximated by
a configuration in which the Higgs has a large expectation value
and its inhomogeneous modes are in vacuum.
This means a drastic simplification in comparison with the scenario
proposed in Refs.~\cite{Krauss:1999ng,Garcia-Bellido:1999sv}
and is obviously only
a first approximation, but it allows us to parameterize the
properties of the inflaton by a single number $\phi_0$, the Higgs
expectation value after the preheating, and therefore makes the
analysis simpler.
In any given inflationary model, it is a straightforward task to
include the inflaton as a dynamical field, thus making the treatment
much more accurate, but at the expense of making the results specific to 
that model.

\section{CP violation}
\label{sect:cp}
In the absence of CP violation, there is no preferred direction for
the change of the baryon number, and therefore domains of matter and
antimatter would be formed. In practice, these domains are far too
small to explain the baryon asymmetry of the universe,
and therefore CP violation is necessary.
Unfortunately,
the only experimentally confirmed source of CP violation
in the standard model,
due to Cabibbo-Kobayashi-Maskawa mixing of quarks,
is too small to cause a sufficient baryon
asymmetry \cite{Shaposhnikov:1987tw,shaposhnikov}.
This means that we must resort to extentions of the
standard model to find additional sources of
C and CP violation.

Typically, effects of CP violation are modelled through an effective field
theory approach.
Once all degrees of freedom
except the gauge fields, the Higgs, and the inflaton have been integrated
out,
the effective Lagrangian will contain non-renormalizable operators
that
break CP. The lowest of these is a dimension-six operator ~\cite{shaposhnikov}
\begin{equation}
\Delta{\cal L}_{\rm CP}
= {\delta_{\rm cp}\over M_{\rm new}^2}\phi^\dagger\phi \,
{3g^2\over32\pi^2}\epsilon^{\mu\nu\rho\sigma}
\,{\rm Tr}\,W_{\mu\nu}W_{\rho\sigma} \,.
\end{equation}
where the dimensionless parameter $\delta_{\rm cp}$ is an
effective measure of CP violation, and $M_{\rm new}$ represents
the scale at which the new physics, responsible for this effective
operator, is important. 
Assuming $M_{\rm new}\sim 1$~TeV,
the experimental constraints on the
magnitude of the neutron electric dipole moment~\cite{Harris99}
lead to an upper bound $\delta_{\rm CP}\lsim 1$, for the CP
violating parameter~\cite{Lue:1997pr}. Recent experimental constraints 
emerging from measurements of the electric dipole moment associated 
with mercury, indicate that the bound on $\delta_{\rm CP}$ could be 
much tighter~\cite{mercury}, being reduced to $\delta_{CP} < 0.05$. 
Such a 
strong bound has to be taken with a degree of caution as its derivation 
assumes a number of things about the magnitude of 
the contributing sources.

There are two ways that we know of to calculate the baryon
number when CP violation is present, each with their own
problems. Firstly one could perform numerical simulations
with an explicit CP violating term in the equations of motions,
as was done in \cite{Garcia-Bellido:1999sv}. However, as
explained in \cite{moore96}, such terms also break the lattice
gauge invariance which is a rather high price to pay. Although
not an insurmountable problem \cite{moore96} we shall not
tackle this here. The second approach is to treat $\delta_{\rm CP}$
perturbatively and not evolve the system with the CP violating
term, rather use this evolution as a background.
For this one measures the change in the
Chern-Simons number
and relates this, via a Boltzmann equation, to the change in the
baryon number. This approach also does not work here as it
relies on $N_{\rm CS}$ following a random walk which, as we shall
argue later, is not the case here, so we do not attempt to measure
the baryon asymmetry in this paper.

\section{Approximations}
\label{sect:approx}
Given the non-perturbative and non-equilibrium nature of the
scenario, it is obviously not possible to study it in the full
standard model without any approximation.
Fortunately,
it is possible to justify three approximations, which together
make the problem tractable:
\begin{itemize}
\item{{\bf Bosonic approximation:}
Initially, most of the energy is in the Higgs field, and
the parametric resonance will transfer it very effectively
into the SU(2) and U(1) gauge fields.
Because a similar parametric resonance cannot take place for the fermionic
fields and because the Higgs couples relatively weakly to most of them,
we assume that the fermions remain in vacuum.
In that case, the main effect of the fermions is to give an extra
channel into which the bosons can decay. We approximate this by adding
a simple damping term with
the damping rate $\gamma\approx 2$~GeV, which agrees with the
experimental decay with of W and Z bosons. This approximation does not
take into account relativistic effects or the oscillations of the
Higgs condensate, but we believe that it gives the correct qualitative
picture. Improved accuracy could be obtained by introducing extra
degrees of freedom that would mimic the effect of the fermions, 
but at the price of extra computational cost.

Of course, these decays also transfer energy into the fermionic sector,
which makes our approximation break down at
$t\gsim \gamma^{-1}$.
However, this does not restrict our analysis much, since if the
transition to the Higgs phase has not taken place by this time,
the reheat temperature will be close to the critical temperature,
and therefore the baryon number is likely to be washed out.
}
\item{{\bf SU(2)$\times$U(1) approximation:}
The Higgs and the SU(2)$\times$U(1) gauge fields interact with SU(3) gluons
only indirectly via fermions. Therefore, if we assume that the fermions
are in vacuum, the same will apply to the gluons as well. In fact,
the flow of energy from fermions to gluons decreases the energy
in the fermionic sector, and therefore makes the
bosonic approximation remain valid for longer times.}
\item{{\bf Classical approximation:}
After preheating, the long-wavelength modes of the Higgs and the gauge
fields have a large occupation number, and therefore they can be treated
as classical fields~\cite{khlebnikov96,Prokopec:1997rr}.
Consequently, time evolution of the system
is given by the classical equations of motion, which can be solved
numerically in a straightforward way.}
\end{itemize}

As a result of these approximations, the time evolution of the system
is given in the temporal gauge
by the classical equations of motion
\begin{eqnarray}
\partial^2_0\phi &=& D_iD_i\phi
+2\lambda \left(\eta^2-\phi^\dagger\phi\right)\phi-\gamma\partial_0\phi,
\nonumber\\
\partial_0^2 B_i&=&-\partial_jB_{ij}+g'{\rm Im}\phi^\dagger D_i\phi
-\gamma\partial_0B_i,
\nonumber\\
\partial_0^2 W_i&=&-[D_j,W_{ij}]+ig
\left[
\phi(D_i\phi)^\dagger
-\frac{1}{2}(D_i\phi)^\dagger\phi-{\rm h.c.}
\right]
-\gamma\partial_0W_i,
\end{eqnarray}
where $\phi$ is the Higgs field, $B_i$ is the U(1) gauge field,
$W_i$ is the SU(2) gauge field, and the covariant derivative is
\begin{equation}
D_i=\partial_i-\frac{i}{2}gW_i-\frac{i}{2}g'B_i.
\end{equation}

Assuming that the decay rate $\gamma$ is the same for all the fields
means that Gauss's law will be conserved. Thus it gives a useful check
for the numerics of the simulation. If we define the canonical momenta
\begin{equation}
\pi=\partial_0\phi,\quad
E_i=-\partial_0B_i,\quad
F_i=-\partial_0W_i,
\end{equation}
Gauss's law has the form
\begin{eqnarray}
\label{equ:gauss}
\partial_iE_i&=&g'{\rm Im}\pi^\dagger\phi,\nonumber\\
\left[D_i,F_i\right]&=&
ig\left[
\pi\phi^\dagger-\frac{1}{2}\phi^\dagger\pi-{\rm h.c.}
\right].
\end{eqnarray}

The temporal gauge leaves
the system invariant under time-independent gauge transformations.
This remaining gauge degree of freedom is fixed by choosing that
the Higgs field is initially real and positive and that only its
lower component has a non-zero value.

Even though we use classical equations of motion, it is essential
to include the quantum vacuum fluctuations in the initial
configuration,
because they act as seeds for the amplification of the low-momentum
gauge field modes~\cite{khlebnikov96}.
They
are approximated by classical fluctuations with the same
equal-time two-point
correlation as in the quantum theory at tree level.
For each real field component $Q$ of mass $m$ and canonical momentum
$P$, this means
\begin{eqnarray}
\label{equ:vaccorr}
\langle Q^*(t,\vec{k})Q(t,\vec{k}')\rangle
&=&
\frac{1}{2\sqrt{\vec{k}^2+m^2}}(2\pi)^3\delta^{(3)}(\vec{k}-\vec{k}'),
\nonumber\\
\langle P^*(t,\vec{k})P(t,\vec{k}')\rangle
&=&
\frac{\sqrt{\vec{k}^2+m^2}}{2}(2\pi)^3\delta^{(3)}(\vec{k}-\vec{k}').
\end{eqnarray}
Since the Higgs is a complex field, it may be decomposed into
four real components
\begin{equation}
\phi=\frac{1}{\sqrt{2}}\left(
\begin{array}{c}
\phi_2+i\phi_3\\
\phi_0+i\phi_1
\end{array}\right).
\end{equation}
Note that as $\phi_0$ has a large value so the masses of these components
are different.

We must also require that the initial configuration satisfies
Gauss's law~(\ref{equ:gauss}).
This is achieved by a cooling procedure
described in Appendix~\ref{app:gauss}.

\section{Lattice}
\label{sect:lattice}
In order to simulate the system numerically, we
discretize it on a $N^3$ lattice with lattice spacing $a$
in such a way that the Higgs field $\phi$
is defined at lattice sites and the gauge fields are represented by
group elements at links between the sites.
In the time direction, the time step is $a_t$ and
the canonical momenta
of the fields are defined at half-way between the time steps.
The details of the discretization, as well as the lattice
equations of motion, are given in Appendix~\ref{app:lattice}.

On a lattice, we label the lattice sites by integer labels
$\chi_j\in \{1,\ldots, N\}$ and the corresponding lattice momenta by
$\kappa_j\in \{0,\ldots N\!-\!1\}$. We define the discrete Fourier transform
by
\begin{eqnarray}
Q(\vec{\kappa})&=&
\frac{1}{N^3}\sum_{\chi_j}\exp\left(2\pi
i\sum_j\chi_j\kappa_j/N\right)Q(\vec{\chi}).
\end{eqnarray}
The physical momentum vector is then given by
\begin{eqnarray}
i\hat{k}_j&=&\frac{1}{a}\left[1-\exp(-2\pi i \kappa_j/N)\right]
\end{eqnarray}
which has the property that the backward lattice derivative acting on
a field in $x$ space is equivalent to multiplying the $\kappa$ space field
by $i\hat{k}_j$. The lattice Laplacian acting in $x$ space
also has the effect in $\kappa$ space of multiplying by
$-\sum_j|\hat{k}_j|^2$, in analogy with the continuum case.

The fields on the lattice are therefore required to satisfy
[cf.~Eq.~(\ref{equ:vaccorr})]
\begin{eqnarray}
\label{equ:latticeic}
\langle Q_{\vec{\kappa}}^*Q_{\vec{\kappa}'}\rangle&=&
\frac{1}{2\sqrt{\sum_j|\hat{k}_j|^2+m_\phi^2}}\frac{1}{N^3 a^3}
\delta_{\vec{\kappa},\vec{\kappa}'}\nonumber\\
\langle P_{\vec{\kappa}}^*P_{\vec{\kappa}'}\rangle&=&
\frac{\sqrt{\sum_j|\hat{k}_j|^2+m_\phi^2}}{2}\frac{1}{N^3 a^3}
\delta_{\vec{\kappa},\vec{\kappa}'}.
\end{eqnarray}
The lattice discretization regularizes the theory,
and it must also be renormalized,
i.e.~the bare parameters used in the lattice equations of motion
must be chosen in such a way that the values of
any observables agree with those measured in experiments.
To our accuracy, we can neglect the renormalization of the
coupling constants, and therefore we only consider the Higgs mass
counterterm. It can be determined by calculating the Higgs mass
to one-loop order in lattice perturbation theory
\begin{equation}
\label{equ:counter}
m_H^2=m_{\rm latt}^2
+\left(6\lambda
+\frac{9}{4}g^2
+\frac{3}{4}g'^2 \right) \langle Q^*Q\rangle_0,
\end{equation}
where the subscript $0$ in $\langle\ldots\rangle_0$
indicates a free-field expectation value,
and $m_H^2$ is the physical Higgs mass.
In the approximation where the masses inside the loops are taken to be
zero, the correlation function (\ref{equ:latticeic}) implies
\begin{equation}
\langle Q^* Q\rangle_0\simeq
\sum_{\kappa_j}\frac{1}{(Na)^3}\frac{1}{2\sqrt{\sum_j|\hat{k}_j|^2}}
\simeq\frac{1}{4\pi^3a^2}\int_0^{\pi}{\rm
d}^3w\left(\sum_j\sin^2 w_j\right)^{-\frac{1}{2}}
\simeq\frac{0.226}{a^2}.
\end{equation}
and $m_H^2$ is the physical Higgs mass. 
This gives the bare mass parameter that
must be used in the simulation,
\begin{equation}
m_{\rm latt}^2\approx m_H^2-\left(6\lambda
+\frac{9}{4}g^2
+\frac{3}{4}g'^2 \right)\frac{0.226}{a^2}.
\end{equation}

\begin{figure}
\center
\epsfig{file=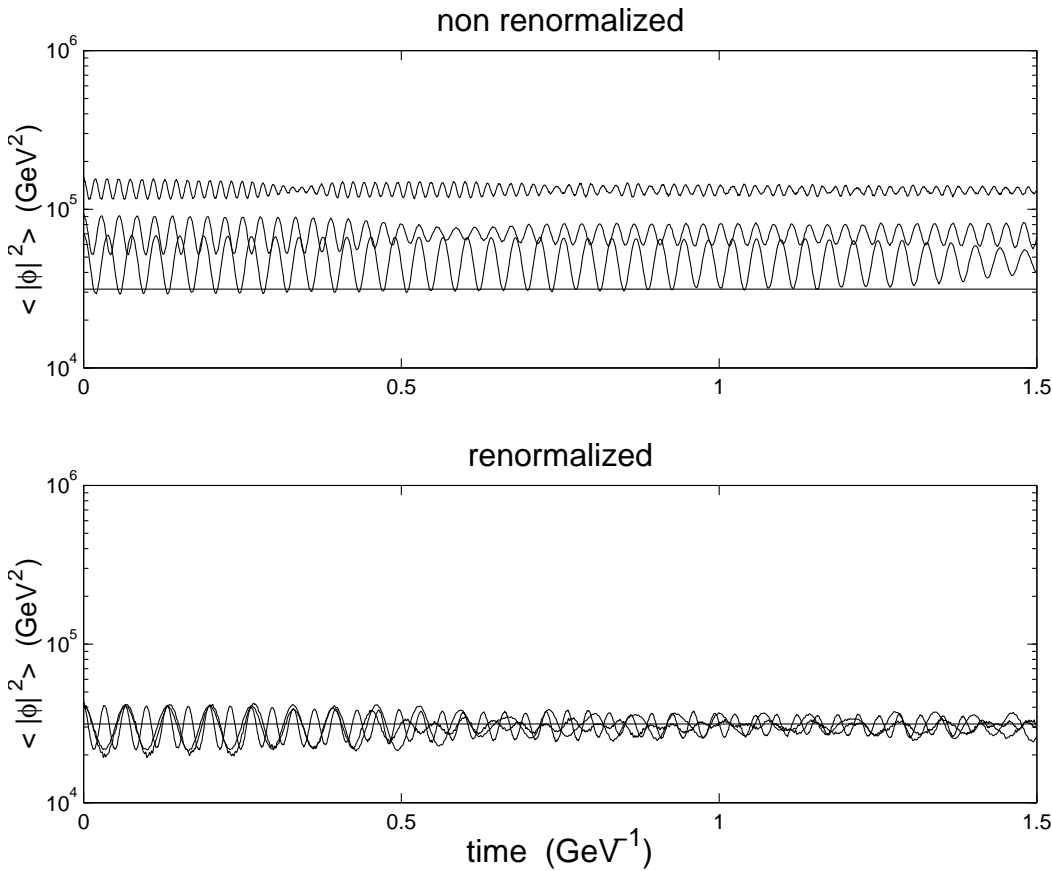,width=14cm}
\flushleft
\caption{
\label{fig:a_dep}
Evolution near the vacuum for a renormalized (bottom) and
non renormalized (top) mass term with $g=g'=0.1$ and
$a=$ $2,3,4\times 10^{-3}$~GeV$^{-1}$.
The straight line is $\eta^2$.}
\end{figure}

Also, we define the ``renormalized'' version of
$\langle\phi^\dagger\phi\rangle$ by subtracting the one-loop
divergence
\begin{eqnarray}
\label{equ:phisqcounter}
\langle\phi^\dagger\phi\rangle&=&\langle\phi^\dagger\phi\rangle_{\rm latt}
-2\langle Q^*Q\rangle_0 \approx
\langle\phi^\dagger\phi\rangle_{\rm latt}-
\frac{0.452}{a^2}.
\end{eqnarray}
This will be a useful quantity as an effective order parameter,
but we emphasize that it is not a physical observable.
In Fig. \ref{fig:a_dep}
we start the fields near the vacuum solution and let them evolve.
The upper graph shows how the equilibrium value of
$\langle\phi^\dagger\phi\rangle$ without the mass counterterm
depends on the lattice spacing,
relaxing to larger values as $a$ is decreased.
When the fluctuations are taken into account by the
counterterm, lower graph,
we see that this dependence is removed. Such behaviour is
essential if we are to have results free from lattice
artefacts.

For our simulation we include a phenomenological damping term
to account for the fermions. The effect of this is to damp the
fluctuations on all length scales, this then changes the two point
correlator (\ref{equ:latticeic}) which appears in the mass counterterm
(\ref{equ:counter}). When the counterterm is calculated one is
interested in how the large momentum modes behave. In the limit
\mbox{$k\gg\gamma$} the dispersion relation becomes
\begin{eqnarray}
\label{equ:dampeffect}
\omega_k\simeq i\frac{\gamma}{2}\pm\sqrt{k^2+m^2},
\end{eqnarray}
showing that \mbox{$\phi_k(t)=\exp(i\omega_k t)\phi_k$} decays
as ${\rm e}^{-\gamma t/2}$ and thus the actual mass counterterm
(\ref{equ:counter}) and $\langle\phi^\dagger\phi\rangle$ counterterm
(\ref{equ:phisqcounter}) must be
multiplied by ${\rm e}^{-\gamma t}$.

\begin{figure}
\center 
\epsfig{file=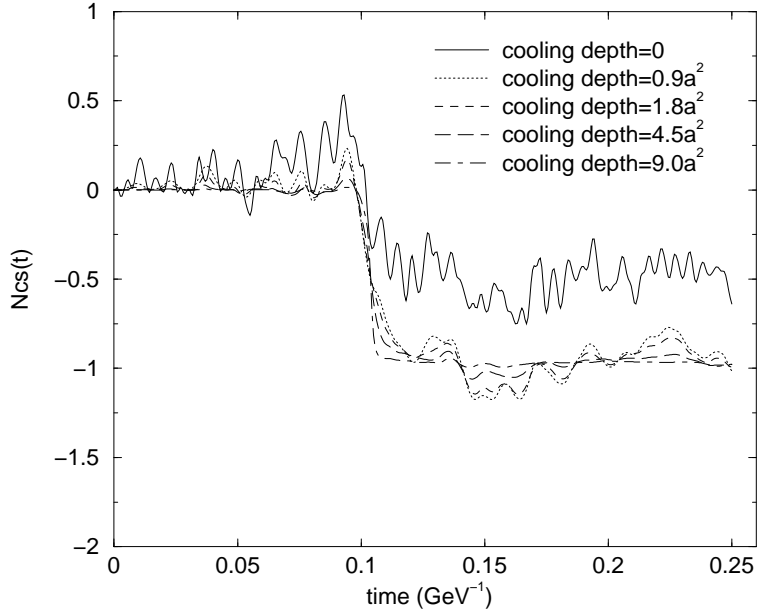,width=10cm} 
\flushleft
\caption{
\label{fig:cool}
Measuring $N_{\rm CS}$ with different cooling paths. The path lengths are 0. 0.9$a^2$,1.8$a^2$, 4.5$a^2$,
9.0$a^2$. Without cooling, the ultraviolet noise blurs the signal, but
with cooling the sphaleron can be seen at $t=0.1~{\rm GeV}^{-1}$.
}
\end{figure}

We also need a lattice representative of the Chern-Simons number
(\ref{cs_number}).
There is no simple way to translate this continuum
expression to a lattice, the essential problem being that any lattice
definition of $\epsilon^{\mu\nu\rho\sigma} W^a_{\mu\nu}
W^a_{\rho\sigma}$ 
is not a total derivative \cite{moore96}.
Here we follow the reasoning of Ambj{\o}rn and Krasnitz \cite{Ambjorn97} where
a daughter set of fields are used to calculate
$\epsilon^{\mu\nu\rho\sigma} W^a_{\mu\nu}
W^a_{\rho\sigma}$. 
In fact, as pointed out in \cite{Moore:1999sw} only the gauge fields should
be included in this daughter set.
This second
set of fields are the cooled image of the first, removing UV
fluctuations and leaving smooth gauge fields, bringing
$\epsilon^{\mu\nu\rho\sigma} W^a_{\mu\nu}
W^a_{\rho\sigma}$ close to a total derivative. This method was developed in
response to the realization that early numerical simulations of the
Chern-Simons diffusion were measuring a large component of thermal
noise rather than actual topological transitions.
To test the method for measuring $N_{\rm CS}$ we evolved the fields in a box
small enough that only one sphaleron could fit in it, thus we should
see only one sphaleron at a time and $N_{\rm CS}$ should change by one unit.
In Fig.~\ref{fig:cool}
we display the difference such a cooling path can make in determining
where a sphaleron transition has taken place. If we measure $N_{\rm CS}$
without using the cooling path (upper curve)
then the UV fluctuations mask the
transition, but as the cooling depth is increased the integer transition
between baryon vacua becomes clearer. In the following we shall use a cooling
depth of $0.9a^2$.

\section{Parameters}
\label{sect:params}
In the simplified model we are discussing here, the
dynamics of the theory depends only on two parameters;
the Higgs mass $m_H$ and the initial
value $\phi_0$ of the Higgs field.
We choose the value $m_H=100$~GeV, but
unlike standard electroweak baryogenesis,
the scenario discussed here is probably not very sensitive to
the Higgs mass.
The coupling constants are taken from electroweak data
\cite{ewdata}; $g=0.64$, $g'=0.34$, $\eta=250.7/\sqrt{2}$~GeV,
$\gamma=2$~GeV.

The initial value $\phi_0$ of the Higgs field parameterizes the
initial energy density, and therefore also determines the final reheat
temperature $T_{\rm rh}$ to which the system eventually equilibrates.
Although in the bosonic theory all the vacua with different $N_{\rm CS}$
are degenerate, the one with zero baryon number has the lowest energy
once the fermions are included
and therefore the sphaleron transitions tend to wash out the baryon
number.
Thus,
if $T_{\rm rh}$ is above or too close to $T_c$, the sphaleron rate
at the final state can be so high that the baryon asymmetry
generated before disappears. It has been estimated that to avoid this,
one needs~\cite{Shaposhnikov:1986jp,Shaposhnikov:1987tw}
\begin{equation}
\label{equ:shapobound}
v(T_{\rm rh})\gsim T_{\rm rh},
\end{equation}
where $v(T_{\rm rh})$ is the expectation value of the Higgs field.
In one-loop perturbation theory,
\cite{kapusta}.
\begin{equation}
v^2(T)=\left(1-\frac{T^2}{T_c^2}\right)2\eta^2,
\quad
T_c^2=\frac{32\lambda \eta^2}{8\lambda+3g^2+g'^2}\approx (200~{\rm GeV})^2,
\end{equation}
and thus the condition (\ref{equ:shapobound})
becomes $T_{\rm rh}\lsim 155$~GeV.
The corresponding energy density is $
\epsilon(T_{\rm rh})\approx (\pi^2/30)g_*T_{\rm rh}^4$,
where $g_*\approx 100$ is the effective number of degrees of freedom
at the electroweak scale.
Since $\epsilon(T_{\rm rh})$ must be equal to the initial energy density
$\epsilon_0=\lambda\phi_0^4$, we find
\begin{equation}
T_{\rm
rh}\approx\left(\frac{30\lambda}{g_*\pi^2}\right)^{1/4}\phi_0
\approx 0.22\phi_0.
\end{equation}
Consequently the constraint (\ref{equ:shapobound}) becomes
\begin{equation}
\label{equ:iniconst}
\phi_0\lsim 700~{\rm GeV}.
\end{equation}

\section{Symmetry restoration}
\label{sect:symmresto} 
\begin{figure}
\center
\epsfig{file=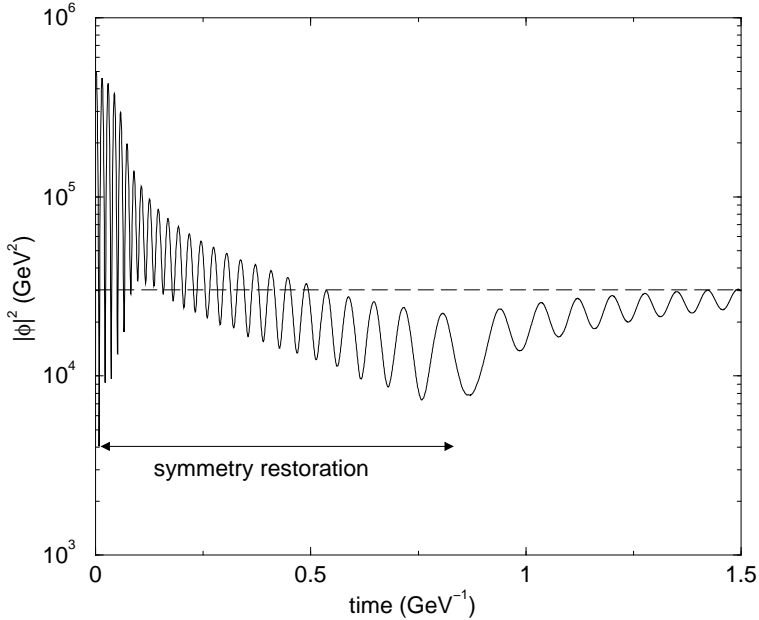,width=10cm}
\flushleft
\caption{
\label{fig:phisq}
The evolution of the renormalised $|\phi|^2$ defined in 
Eq.~(\protect\ref{equ:phisqcounter})
as a function of time. Because of the inhomogeneous fluctuations,
$|\phi|^2$ never vanishes, but its exponential decay shows that 
the electroweak symmetry is effectively restored up until $t\gsim
0.8$~GeV$^{-1}$}
\end{figure}

In Fig.~\ref{fig:phisq} we show the time
evolution of the ``renormalized'' value of $\langle
|\phi|^2\rangle$ [see Eq.~(\ref{equ:phisqcounter})]. We use a
$60^3$ lattice with the initial condition $\phi_0=700~{\rm GeV}$
[cf.~Eq.~(\ref{equ:iniconst})]
and lattice spacings $a=3\times 10^{-3}$~GeV$^{-1}$, $a_t=0.2a$;
unless otherwise stated these will be used throughout the paper.
We also performed test runs with smaller lattice spacings and time
steps, 
finding
results that were in statistical agreement with those presented here.

Since $\langle|\phi|^2\rangle$ gets a positive
definite contribution from every Fourier mode, it is not zero even in
the symmetric phase. Furthermore, since there is no phase transition
in the electroweak theory even in equilibrium~\cite{Kajantie:1997qd},
there is no rigorous way to determine whether the electroweak
symmetry is restored non-thermally or not. Nevertheless, we know that
the damping term causes the amplitude of the inhomogeneous modes to
decrease according to Eq.~(\ref{equ:dampeffect}), and thus
if there is no Higgs condensate present we expect
$|\phi|^2$ to decrease as $\exp(-\gamma t)$. This is indeed what we
observe
until $t\approx 0.8$~GeV$^{-1}$. We can therefore conclude that the
electroweak symmetry is effectively restored during this time.

At $t\gsim
0.8$~GeV$^{-1}$, $\langle|\phi|^2\rangle$ starts to grow and approaches its vacuum
expectation value. Although it is not entirely clear that our
approximations are still valid at this time, because
$t\gsim\gamma^{-1}$, conservation of energy implies that the system
must
eventually end up in the broken phase, because the initial energy
density is lower than that at the transition point in equilibrium.

In order to obtain a better insight to the dynamics of the fields
we also measured the power spectrum of the electric field
$E_i$ of the U(1) gauge field $B_{\mu}$.
It is a gauge-invariant quantity, and although Gauss's
law~Eq.~(\ref{equ:gauss}) fixes the value of its longitudinal component,
it does not affect its transverse
components. In thermal equilibrium at temperature $T$, each Fourier
mode
$E^{\rm T}_i(k)$ of the transverse components satisfies
\begin{equation}
\frac{d^3k}{(2\pi)^3}\langle|E^{\rm T}_{i}(k)
|^2\rangle=2T,
\end{equation}
and therefore one can define even in a non-equilibrium setting an
effective
temperature for each Fourier mode separately~\cite{rajantie00}
\begin{equation}
\label{equ:Teff}
T_{\rm eff}(k)=\frac{1}{2}|E^{\rm T}_{i}(k)
|^2\frac{d^3k}{(2\pi)^3}.
\end{equation}

\begin{figure}
\center
\epsfig{file=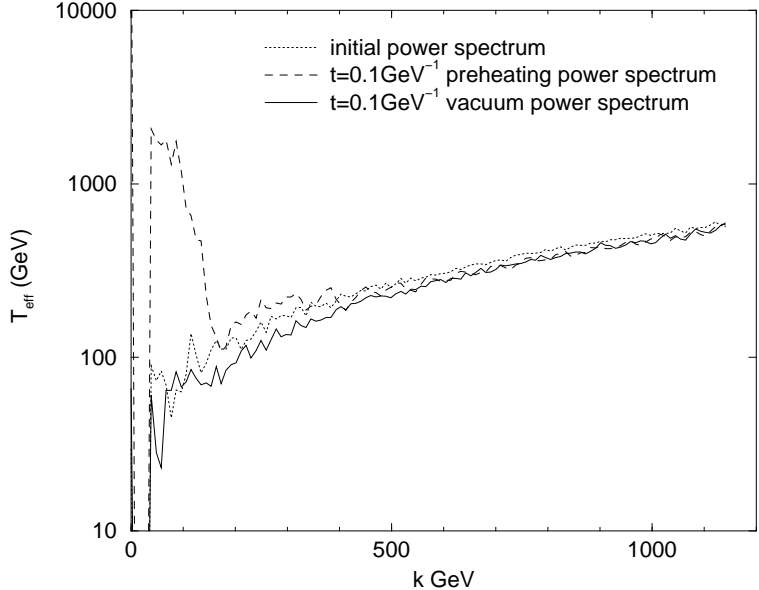 ,width=10cm}
\flushleft
\caption{
\label{fig:Teff}
Effective temperature (\protect\ref{equ:Teff}) 
of different Fourier modes of the hypercharge field measured in the
initial configuration and at $t=0.1~{\rm GeV}^{-1}$. 
At this time, the Higgs fluctuations have heated up the
long-wavelength modes to a high effective temperature.
For comparison, the plot also shows the effect of the damping term to
the vacuum,
measured in a run with $\phi_0=0$.}
\end{figure}

In Fig.~\ref{fig:Teff}, we show the effective temperature of different
Fourier modes of $E_i$ at the beginning of the simulation and at time
$t=0.1$~GeV$^{-1}$. Note that although the electric field is initially
in vacuum, its effective temperature is nonzero, $T_{\rm
eff}(k)\approx k/2$.
This is caused by the initial ``quantum'' fluctuations
(\ref{equ:vaccorr}),
and corresponds to the vacuum energy $\omega/2$ of a quantum harmonic
oscillator.
In fact, in a quantum system, the effective temperature
would be related to the occupation number as
\begin{equation}
n(k)=\frac{T_{\rm eff}(k)}{k}-\frac{1}{2}.
\end{equation}

At $t=0.12$~GeV$^{-1}$, the long-wavelength modes with
$k\lsim k_*\approx200$~GeV
have reached a high
temperature $T_{\rm eff}\approx 10^4$~GeV, but the short-wavelength
modes are still in vacuum. (The evolution of the power spectrum at
large $k$ is due to the damping factor, but these modes remain
vacuum modes as the counterterm is also decaying.)

This form of the power spectrum justifies the classical approximation,
because the occupation number of the long-wavelength modes is high,
$n\sim 100$, and they are the relevant modes for the time evolution.
The main effect of the short-wavelength modes is to renormalize the
couplings, and that has been taken into account
perturbatively~(\ref{equ:counter}).

As argued in Ref.~\cite{rajantie00},
the cutoff momentum is parametrically of order
$k_*\sim g\phi_0$, and the effective temperature
$T_{\rm eff}\sim\phi_0/g$. This is compatible with
the observation \cite{Linde:1980ts} that the theory becomes non-perturbative
at momenta
$k\lsim g^2T$, and thus the interactions below $k_*$ are
not suppressed by the coupling constants and thermalization
is more effective.

Eventually, $k_*$
moves towards higher momenta, but this process is relatively
slow~\cite{rajantie00}. Instead, the thermalization is dominated by the
decays of the long-wavelength bosons into fermions, which is
approximated by the damping term.

Obviously, it would be more interesting to follow the evolution of
the power spectrum of the SU(2) field, but because its electric field
is not gauge invariant, this is not possible.
There are indications \cite{Guy} that in the ``vacuum'' configuration
(\ref{equ:latticeic}),  the cascade of power from the
ultraviolet to the infrared modes is significantly faster in the SU(2)
field than in scalar or U(1) fields. This would lower the ultraviolet
part and raise the infrared part of power spectrum in
Fig.~\ref{fig:Teff}. However, since the power spectrum
is already dominated by the infrared modes, this extra contribution to
the infrared is not likely to be significant. Likewise, once the
infrared modes have been excited, their occupation numbers are so high
that the precise form of the ultraviolet
power spectrum is unimportant.

It is also possible that the energy transfer to the opposite
direction, from infrared to ultraviolet, is faster in the SU(2)
case, which would mean that the whole SU(2) field could thermalize much
faster than the U(1) field in Fig.~\ref{fig:Teff}.
Even if this is the case, it would not
presumably change the qualitative behaviour, because the SU(2) gauge
field only constitutes a small fraction of the total field content of
the standard model; full thermal
equilibrium is not reached before all the fermions have equilibrated
as well. Therefore we believe that our classical approximation gives
the correct qualitative picture of the dynamics.

During the period of symmetry restoration, baryon number is not
conserved, and since the effective temperature of the long-wavelength
modes
is $T_{\rm eff}\approx 10^3$~GeV, we can estimate using
Eq.~(\ref{equ:sphest}) that the sphaleron rate is roughly
\begin{equation}
\label{equ:gammapred}
\Gamma_{\rm
sph}\approx 10^6~{\rm GeV}^4. 
\end{equation}
Thus the non-equilibrium 
oscillations ought to be able to
generate baryon asymmetry \cite{Garcia-Bellido:1999sv}.
On the other hand, the rate at which the temperature decreases,
$\gamma\approx 2$~GeV, is much higher than the baryon decay rate
$\gamma_B\approx 0.01\ldots0.1$~GeV in Eq.~(\ref{equ:gammab}),
which means that the sphaleron rate drops to zero so rapidly that
the baryon asymmetry should not be washed out.

\section{Evolution of the Chern-Simons number}
\label{sect:Ncs}
In thermal equilibrium, the sphaleron rate $\Gamma_{\rm sph}$ can be
measured directly, because
$N_{\rm CS}$ performs a random walk and
$\Gamma_{\rm sph}$ is given by the diffusion
constant
\begin{equation}
\label{equ:Gammadef}
\Gamma_{\rm sph}=\lim_{t\rightarrow\infty}
\frac{
\langle \left(N_{\rm CS}(t) - N_{\rm CS}(0)\right)^2
\rangle}{t}.
\end{equation}
In Ref.~\cite{Garcia-Bellido:1999sv}, this was generalized to
non-equilibrium cases by defining the time-dependent sphaleron rate
\begin{equation}
\label{equ:gammat}
\Gamma_{\rm sph}(t)=\frac{d}{dt}
\langle \left(N_{\rm CS}(t) - N_{\rm CS}(0)\right)^2
\rangle.
\end{equation}
This was then used in a phenomenological Boltzmann equation to
estimate the baryon asymmetry generated during preheating.

\begin{figure}
\center
\epsfig{file=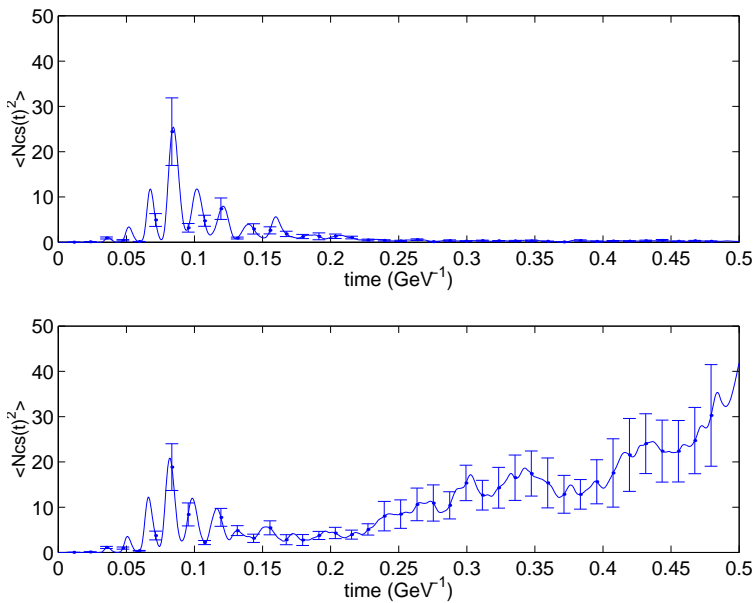 ,width=10cm}
\flushleft
\caption{
\label{fig:Ncs_evolve}
$\langle N_{\rm CS}^2\rangle$ for dissipation
$\gamma=2$~GeV$^{-1}$ (top) and $\gamma=0$ (bottom)
with a cooling depth of $0.9a^2$, averaged over 12 runs.
Initially $\langle N_{\rm CS}^2\rangle$ oscillates and then starts to
decrease vanishing eventually, which means that no topological
transitions actually took place.
In thermal equilibrium, $\langle N_{\rm CS}^2\rangle$ would grow
linearly, which is what happens in the dissipationless run at late
times.
}
\end{figure}

In Fig.~\ref{fig:Ncs_evolve}, we show the time evolution of the
variance, $\langle \left(N_{\rm CS}(t) - N_{\rm CS}(0)\right)^2
\rangle$, of the Chern-Simons number during the simulation (upper
plot). Instead of growing monotonically, as Eq.~(\ref{equ:gammat})
assumes, the variance oscillates until $t\approx 0.15~$~GeV$^{-1}$,
and then finally settles down to a value that is well below its
maximum during the oscillations. This indicates that the behaviour
of $N_{\rm CS}$ is not purely Brownian, but it contains a
deterministic component, which tries to decrease its value. To see
if this is a result of the damping we performed simulations where
$\gamma=0$. The lower graph in Fig.~\ref{fig:Ncs_evolve} shows
that the oscillations persist even if we remove the damping term.
In that case, $N_{\rm CS}$ keeps on wandering after the initial
oscillations because the system thermalizes to a high temperature,
but the behaviour is still non-Brownian at $t\approx
0.1$~GeV$^{-1}$. In fact we would not expect the damping to have
an effect as the damping timescale is $0.5$~GeV$^{-1}$ and the
effect occurs before $0.1$~GeV$^{-1}$.

This deterministic behaviour is explained by the interaction between
the gauge and Higgs fields. When the Higgs field $\Phi$ is
non-vanishing, one can define the integer-valued Higgs
winding number $N_{H}$ \cite{turok90},
\begin{equation}
N_H=-\frac{1}{24\pi^2}\int d^3x\epsilon^{ijk}
{\rm Tr}\left(\partial_i\Phi^\dagger\partial_j\Phi
\partial_k\Phi^\dagger\Phi\right).
\end{equation}
In particular, in a vacuum $N_{\rm CS}=N_{H}$,
because any vacuum configuration can be gauge transformed into one in
which the gauge field vanishes, which implies
$N_{\rm CS}=0$, and the Higgs field is constant, which implies $N_H=0$.

In previous studies of $N_{\rm CS}$ in thermal 
equilibrium~\cite{Ambjorn90,Ambjorn:1995xm,Moore:1997bn,%
Moore:1998sn,%
Moore:2000fs,Bodeker:2000gx}, 
this has not been a problem. 
If we write
\begin{equation}
N_{\rm CS}=N_++N_-,\quad{\rm where}\quad
N_+=\frac{N_{\rm CS}+N_H}{2},\quad
N_-=\frac{N_{\rm CS}-N_H}{2},
\end{equation}
the different vacua are labelled by $N_+$. 
In thermal equilibrium, $N_+$ performs a random walk because of
sphaleron transitions, and
$N_-$ merely fluctuates around zero. In our non-equilibrium case,
these
fluctuations must die away as the system cools and
therefore $N_-$ must tend to zero. Thus,
$N_+$ will
be the sole contributor to $N_{\rm CS}$ in the final state.
Given that, the contribution to $\langle N_{\rm CS}^2\rangle$ 
from $N_+$ will be
monotonic and all the early oscillations of Fig.~\ref{fig:Ncs_evolve}
are due to $N_-$.
This implies that practically no topological sphaleron transitions
took place in any of our runs, while Eq.~(\ref{equ:gammapred}) would
have
predicted around 5000 of them.

\begin{figure}
\center
\epsfig{file=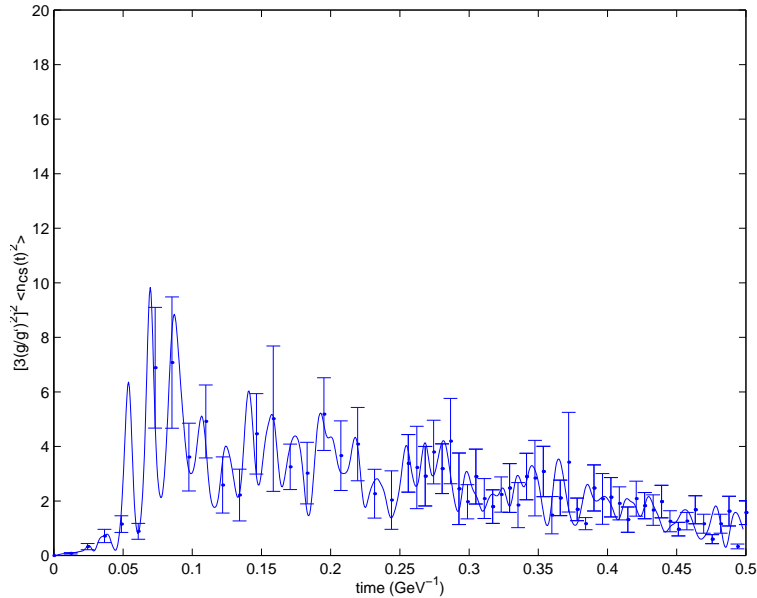 ,width=10cm}
\flushleft
\caption{
\label{fig:ncs_evolve}
The squared U(1) Chern-Simons number $\langle n_{\rm CS}^2\rangle$
multiplied by the factor
$\left[3(g/g')^2\right]^2$ [see Eq.~(\protect\ref{equ:ncsratio})]. It is of
similar order of magnitude as $\langle N_{\rm CS}^2\rangle$ in
Fig.~\protect\ref{fig:Ncs_evolve}, which supports the idea that $\langle
N_{\rm CS}^2\rangle$ consists of fluctuations only.
}
\end{figure}

We can further back up this claim that the changes in 
$\langle N_{\rm CS}^2 \rangle$ are not of a topological
nature by
looking at $\langle n_{\rm CS}^2 \rangle$, 
the U(1) equivalent of $\langle N_{\rm CS}^2 \rangle$,
[see Eq.~(\ref{ncs_number})];
large gauge transformations leave $\langle n_{\rm CS}^2 \rangle$
unaffected and so its evolution is due only to fluctuations.
If there is no topology change in the SU(2) sector then we can
relate the fluctuations of $\langle N_{\rm CS}^2 \rangle$ to
$\langle n_{\rm CS}^2 \rangle$ via \cite{Ambjorn:1995xm}%
\begin{eqnarray}
\label{equ:ncsratio}
\langle N_{\rm CS}^2 \rangle/\langle n_{\rm CS}^2 \rangle
\simeq\left[3(g/g')^2\right]^2,
\end{eqnarray}
where the factor 3 is  because there are
three SU(2) vector bosons to the one U(1) vector. The plot of
$~\left[3(g/g')^2\right]^2\langle n_{\rm CS}^2\rangle$ in 
Fig. \ref{fig:ncs_evolve} shows that the U(1) Chern-Simons number
is of the order expected if the evolution does not contain topological
changes.

This non-Brownian element in the motion of $N_{\rm CS}$
means that the simple Boltzmann equation does not give an
adequate description of the dynamics. The same effect is present also
in the CP violating case, and this may lead to baryon washout even
well below $T_c$ if the Higgs winding remains small during the
non-equilibrium time evolution. Even if the initial fluctuations generate
large $N_{\rm CS}$, and therefore a large baryon asymmetry, the
interaction with the Higgs field may destroy it soon afterwards.

\section{Conclusions}
\label{sect:concl}

In this paper, we have studied the out-of-equilibrium dynamics of the
electroweak theory in a case where most of the energy is concentrated
in the long-wavelength modes.
These initial conditions are interesting because they resemble the
state after preheating at the electroweak scale,
where the Higgs has a large
initial value. All the other degrees of freedom were in vacuum, and
were given initial conditions that reproduce the tree-level two-point
function of the quantum vacuum. The coupling to the fermions was
approximated by a phenomenological damping term, whose effects were
also taken into account in the renormalization by using a
time-dependent mass counterterm.

Our method can be used to simulate more realistic scenarios of
electroweak preheating in a straightforward way by coupling the Higgs
to the inflaton field. 
In order to achieve accurate results, one would also have to be able
to improve on our approximations, but we believe that even in its present
form, the method can give reliable order-of-magnitude estimates.

Our results show that the oscillations of the Higgs field can lead to
non-thermal restoration of the SU(2) gauge symmetry.
This supports the scenario of baryogenesis where the out
of equilibrium state, required by the Sakharov criteria,
is provided by a form of parametric resonance rather than
a thermal phase transition
\cite{Krauss:1999ng,Garcia-Bellido:1999sv}, but
a truly quantitative test would require a measurement of the
generated baryon asymmetry. In our simulations this was not possible
because the equations of motion were CP invariant.

However, we also
studied the Chern-Simons number of the SU(2) gauge field and
showed that its motion is not pure random walk. We believe this
is
due to interactions with the Higgs fields which force the vacuum
value of the Chern-Simons number to equal the winding number of
the Higgs field. If the Higgs winding therefore does not change
significantly, these interactions will reduce $N_{\rm CS}$, destroying
the baryons previously created. Conversely, if a mechanism greatly
enhanced $N_H$ then baryons would be created as the winding
numbers equilibrated.
In every one of our simulations, the final Chern-Simons number
vanished, which suggests that the rate of true topological transitions
that would lead to stable baryons is lower than expected during the
non-thermal symmetry restoration. It remains to be seen if this is
true in more realistic models as well.
In any event, it means that the baryon asymmetry cannot be estimated
using any effective description that does not take this effect into account.

%%%%%%%%%%%%%%%%%%%%%%%%%%%%%%%%%%%%%%%%%%%%%%%%%%%%%%%%
\acknowledgements We would like to thank D. B\"odeker for useful
conversations
and G.D.~Moore for informing us about his results~\cite{Guy}. 
EJC, AR and PMS are supported by PPARC, and AR also
partly by the University of Helsinki. This work was conducted on
the SGI Origin platform using COSMOS Consortium facilities, funded
by HEFCE, PPARC and SGI.

%%%%%%%%%%%%%%%%%%%%%%%%%%%%%%%%%%%%%%%%%%%%%%%%%%%%%%%%
\appendix
\section{Evolution equations}
\label{app:lattice}
To be explicit we give here the lattice method~\cite{Ambjorn90,Moore:1999sw} 
used to simulate the
continuum gauge theory. The starting point is the gauge covariant
derivative on a lattice,
\begin{eqnarray}
\tilde{D}_i\phi&=&\frac{1}{a}\left(U_i(t,x)V_i(t,x)
\phi(t,\xi)-\phi(t,x)\right),\\
\tilde{D}_0\phi&=&\frac{1}{a_t}
\left(U_0(t,x)V_0(t,x)\phi(t+a_t,x)-\phi(t,x)\right).
\end{eqnarray}
Here the $\tilde{}$ refers to the lattice covariant derivative, and $a$
and $a_t$ are the lattice spacings in the space and time directions,
respectively, with
$\xi$ referring to the position one lattice site away in the
$i$ direction. The fields $U_i(t,x)$ and $V_i(t,x)$ are fundamental
SU(2), U(1) valued respectively, and correspond to the continuum
gauge fields according to,
\begin{eqnarray}
U_i(t,x)&=&\exp\left(-\frac{i}{2}g a W^a_i \sigma^a \right),\\
U_0(t,x)&=&\exp\left(-\frac{i}{2}g a_t W^a_0 \sigma^a \right),\\
V_i(t,x)&=&\exp\left(-\frac{i}{2}g' a B_i \right),\\
V_0(t,x)&=&\exp\left(-\frac{i}{2}g' a_t B_0 \right).
\end{eqnarray}
One then finds that the standard continuum gauge action is modelled
by
\begin{eqnarray}
\tilde{S}&=&\sum_{x,t}a_t a^3 \left[
 \tilde{D}_0\phi^\dagger\tilde{D}_0\phi
-\sum_i \tilde{D}_i\phi^\dagger\tilde{D}_i\phi
-{\cal V}(|\phi |)\right.\\
&~&\;\;\;\;\;\;\;\;\;\;\;\;\;\;
\nonumber-\frac{2}{a_t^2 a^2 g^2}  \sum_i {\rm Tr}\left(U_{0i}\right)
-\frac{4}{a_t^2 a^2 g'^2} \sum_i {\rm Tr}\left(V_{0i}\right)\\
\nonumber &~&\;\;\;\;\;\;\;\;\;\;\;\;\;\;\left.
+\frac{1}{a^4 g^2}  \sum_{ij} {\rm Tr}\left(U_{ij}\right)
+\frac{2}{a^4 g'^2} \sum_{ij} {\rm Tr}\left(V_{ij}\right)\right].
\end{eqnarray}
Here we do {\it not} use the repeated index summation convention.
The field strengths for the gauge fields are given by
\begin{eqnarray}
U_{ij}(t,x)&=&U_j(t,x)U_i(t,\xj)U_j^\dagger (t,\xi)U_i^\dagger(t,x),\\
V_{ij}(t,x)&=&V_j(t,x)V_i(t,\xj)V_j^\dagger(t,\xi) V_i^\dagger(t,x),
\end{eqnarray}
which means that the lattice action has the following gauge symmetry
\begin{eqnarray}
\phi(t,x)&\longrightarrow&\Omega_1(t,x)\Omega_2(t,x)\phi(t,x),\\
U_i(t,x)&\longrightarrow&\Omega_2(t,x)U_i(t,x)\Omega_2^\dagger(t,\xi),\\
V_i(t,x)&\longrightarrow&\Omega_1(t,x)V_i(t,x)\Omega_1^\dagger(t,\xi),\\
\Omega_2&\in& {\rm SU(2)},\;\;\Omega_1\in {\rm U(1)}.
\end{eqnarray}
In our simulations we use the equivalent of temporal gauge, in which
the link variables $U_0$ and $V_0$ assume the identity value, with
their equations of motion leading to Gauss type constraints.
The equations of motion are found by the lattice equivalent of
functional differentiation of the lattice action,
\begin{eqnarray}
\phi(t\!+\!a_t,x)&=&\phi(t,x)+a_t\pi(t\!+\!a_t/2,x),\\
V_i(t\!+\!a_t,x)&=&\frac{1}{2}g'aa_t E_i(t\!+\!a_t/2,x)V_i(t,x),\\
U_i(t\!+\!a_t,x)&=&g aa_t F_i(t\!+\!a_t/2,x)U_i(t,x),\\
\pi(t\!+\!a_t/2,x)&=& (1\!-\!a_t \gamma)\pi(t\!-\!a_t/2,x)\\
\nonumber             &~&+a_t\left[ \frac{1}{a^2}\sum_i\left(
U_i(t,x)V_i(t,x)\phi(t,\xi)-2\phi(t,x)\right.\right.\\
\nonumber
&~&\left.\left.\;\;\;\;\;\;\;\;+U_i^\dagger(t,x)V_i^\dagger(t,x)
\phi(t,x\!-\!\hat{\imath})\right)
-\frac{\partial {\cal V}}{\partial\phi^\dagger}\right],\\
{\rm Im}(E_k(t\!+\!a_t/2,x))&=&(1\!-\!a_t \gamma){\rm Im}(E_k(t\!-\!a_t/2,x))\\
\nonumber
&~&+a_t\left[\frac{g'}{a}{\rm Im}[\phi^\dagger(t,\xk) U_k^\dagger
(t,x)                                      
V_k^\dagger(t,x) \phi(t,x)]\right.\\
\nonumber     &~&\hspace{2em}-\frac{2}{g'a^3}\sum_i
{\rm Im}[V_k(t,x)V_i(t,\xk)
V_k^\dagger(t,\xi)
V_i^\dagger(t,x)\\
\nonumber
&~&\left.\hspace{7em}+V_i(t,x\!-\!\hat{\imath})V_k(t,x)
V_i^\dagger(t,x\!+\!\hat{k}\!-\!\hat{\imath})
V_k^\dagger(t,x\!-\!\hat{\imath}) ]
\right],\\
{\rm Tr}(i\sigma^m F_k(t\!+\!a_t/2,x))&=&
           (1\!-\!a_t \gamma){\rm Tr}(i\sigma^m F_k(t\!-\!a_t/2,x))\\
\nonumber
&~&+a_t\left[\frac{g}{a}{\rm Re}
[\phi^\dagger(t,\xk)U_k^\dagger(t,x)
V_k^\dagger(t,x)i\sigma^m\phi(t,x)]\right.\\
\nonumber         &~&\hspace{1em}-\frac{1}{g a^3}\sum_i 
{\rm Tr}[i\sigma^m
U_k(t,x)U_i(t,\xk)
U_k^\dagger(t,\xi)
U_i^\dagger(t,x)\\
\nonumber&~&\left.\hspace{6em} +i\sigma^m
U_k(t,x)
U_i^\dagger(t,x\!+\!k\!-\!i)
U_k^\dagger(t,x\!-\!i)
U_i(t,x\!-\!i)]\right].
\end{eqnarray}
Here we have included a damping factor $\gamma$ which is a
phenomenological term put in by hand to model the decay into fermions.
Such a term does not spoil the gauge invariance of the model.
The remaining components of $E_k$ and $F_k$ are found by using
\begin{eqnarray}
\det(E)&=&\frac{2}{g'aa_t},\hspace{3em}\det(F)=\frac{1}{gaa_t}.
\end{eqnarray}
The Gauss constraints which are implied by the lattice gauge symmetry
are
\begin{eqnarray}
\label{hypergauss}
\frac{1}{a}\sum_i {\rm Im}[E_i(t\!+\!a_t/2,x)-E_i(t\!+\!a_t/2,x\!-\!\hat{\imath})]\\
\nonumber
-g'{\rm Im}[\pi^\dagger(t\!+\!a_t/2,x)\phi(t,x)]&=&0,\\
\label{weakgauss}
\frac{1}{a}\sum_i {\rm Tr}[ i\sigma^k F_i(t\!+\!a_t/2,x)
-i\sigma^k U_i^\dagger(t,x\!-\!\hat{\imath}) F_i(t\!+\!a_t/2,x\!-\!\hat{\imath})
U_i(t,x\!-\!\hat{\imath})\\
\nonumber
    -g{\rm Re}[\pi^\dagger(t\!+\!a_t/2,x) i\sigma^k \phi(t,x)]&=&0.
\end{eqnarray}

%%%%%%%%%%%%%%%%%%%%%%%%%%%%%%%%%%%%%%%%%%%%%%%%%%%%%%%%
\section{Restoring the Gauss constraints.}
\label{app:gauss}

When the random initial conditions on the gauge and scalar
fields are imposed we find that the Gauss constraints will
generically be violated. In order that the simulation is to
make sense this must be remedied. In the Abelian case the form of
Gauss's law is simple in $k$ space so the longitudinal component of $E_j$
may easily be set to the relevant
charge density. In the non Abelian case however the derivative in Gauss's
law requires a gauge field to make it gauge covariant, it is not such
a simple matter therefore to find the equivalent of the transverse and
longitudinal components. The way we get around this is to relax the
unphysical gauge degrees of freedom in the electric fields and scalar
momenta into a state which satisfies Gauss's law.
This is done by making a Hamiltonian out of the Gauss constraint and
using this to define a dissipative motion 
\cite{Ambjorn90,moore96}.

Defining $G^0$ to be the hypercharge Gauss constraint expression
(\ref{hypergauss}) and $G^k$ to be the remaining three expressions
for the SU(2) field (\ref{weakgauss}), such
that we require $G^0=0$, $G^k=0$. Now
introduce the Hamiltonian
\begin{eqnarray}
H&=&\sum_x a^3\left[G^0 G^0 + G^k G^k \right].
\end{eqnarray}
By writing \mbox{$F_i=F_i^0-\frac{i}{2}\sigma^\alpha F_i^\alpha$}
we then evolve the $\pi$, $E_i$, $F_i$ fields according to the
dissipative equations
\begin{eqnarray}
\frac{\partial \psi}{\partial \tau}&=&-\frac{\delta H}{\delta\psi},
\end{eqnarray}
where $\psi$ represent the generic momentum field, to find
\begin{eqnarray}
\frac{\partial \pi}{\partial\tau}&=&
-\left[ ig'G^0(\tau,x)\phi(\tau,x)
        -ig\sum_k \sigma^k G^k(\tau,x)\phi(\tau,x) \right],\\
\frac{\partial {\rm Im}(E_j)}{\partial\tau}&=&
         -\frac{2}{a}\left[G^0(\tau,x)-G^0(\tau,\xj)\right],\\
\frac{\partial F^\beta_j}{\partial\tau}&=&
         -\frac{2}{a}\left[G^\beta(\tau,x)
                 -\frac{1}{2}{\rm Tr}\left(\sigma^k
U^\dagger_j(\tau,x)\sigma^\beta U_j(\tau,x)\right)G^k(\tau,\xj)\right].
\end{eqnarray}

%%%%%%%%%%%%%%%%%%%%%%%%%%%%%%%%%%%%%%%%%%%%%%%%%%%%%%%%

\end{document}